\documentclass[12pt]{article}
\usepackage[margin=1in, paperwidth=8.5in, paperheight=11in]{geometry}
\usepackage{amsmath}
\usepackage{graphicx}%
\usepackage{amsfonts}%
\usepackage{amssymb}
\usepackage{color}
\usepackage{float,subfigure}
\usepackage[round]{natbib}
\usepackage{cite}
\usepackage{setspace}
\usepackage{comment}
\usepackage{url}
\usepackage{enumerate}
\usepackage{verbatim}
\usepackage{rotating}

\newcommand{\bet}{ \mbox{\boldmath $ \eta $} }

\newcommand{\bbeta}{ \ensuremath{\boldsymbol{\beta}}}

\newcommand{\bphi}{ \mbox{\boldmath $\phi$}}

\newcommand{\Sig}{ \mbox{$\Sigma$} }
\newcommand{\sig}{ \ensuremath{\sigma}}

\newcommand{\btau}{ \mbox{\boldmath $\tau$}}

\newcommand{\brho}{ \mbox{\boldmath $\rho$} }

\newcommand{\iid}{ \mbox{$\stackrel{\text{iid}}{\sim}$}}

\newcommand{\bzero}{\textbf{0}}

\newcommand{\bv}{ {\bf v} }

\newcommand{\bx}{ {\bf x} }

\newcommand{\bY}{ {\bf Y} }

\newcommand{\bZ}{ {\bf Z} }

\newcommand{\given}{\,\vert\,}

\newcommand{\tR}{\widetilde{R}}

\newcommand{\MCAR}{\mbox{$\text{MCAR}$}}
\newcommand{\MSTCAR}{\mbox{$\text{MSTCAR}$}}
\newcommand{\STCAR}{\mbox{$\text{STCAR}$}}
\newcommand{\tbeta}{\mbox{$\text{Beta}$}}
\newcommand{\IW}{\mbox{$\text{InvWish}$}}
\newcommand{\Wish}{\mbox{$\text{Wish}$}}
\newcommand{\Pois}{\mbox{$\text{Pois}$}}

\begin{document}

%\author{Harrison Quick, Bradley P.\ Carlin, and Sudipto Banerjee \footnote{Harrison Quick is a doctoral student, Bradley P. Carlin (E-mail:\textit{carli002@umn.edu}) is Mayo Professor and Chair of Biostatistics, and Sudipto Banerjee is Professor of Biostatistics, School of Public Health, Division of Biostatistics, University of Minnesota, Minneapolis, MN 55455.}}

%\title{Heteroscedastic CAR models for areally referenced temporal processes with an application to California asthma hospitalization data}
%\date{\today}
%\maketitle
\thispagestyle{empty}
\setcounter{page}{0}

\begin{center}
{\Large \textbf{{A Nonseparable Multivariate Space-Time Model for Analyzing County-Level Heart Disease Death Rates by Race and Gender}}}

\bigskip

\begin{comment}
\textbf{Harrison Quick}\\ %, Michael R.\ Kramer$^{2}$, Adam S.\ Vaughan$^{2}$, Linda Schieb$^{1}$, and Sophia Greer$^{1}$}\\
Division of Heart Disease and Stroke Prevention, Centers for Disease Control and Prevention, Atlanta, GA 30329\\
%$^3$ Department of Epidemiology, Emory University, 1518 Clifton Rd NE, Atlanta, GA 30322\\
%$^{*}$
\emph{email:} HQuick@cdc.gov
%Last updated: \today
\end{comment}
%\begin{comment}
\textbf{Harrison Quick$^{*1}$, Lance A.\ Waller$^{2}$, and Michele Casper$^1$}\\ %, Michael R.\ Kramer$^{2}$, Adam S.\ Vaughan$^{2}$, Linda Schieb$^{1}$, and Sophia Greer$^{1}$}\\
$^{1}$ Division of Heart Disease and Stroke Prevention, Centers for Disease Control and Prevention, Atlanta, GA 30329\\
$^2$ Department of Biostatistics, Emory University, 1518 Clifton Rd NE, Atlanta, GA 30322\\
%$^3$ Department of Epidemiology, Emory University, 1518 Clifton Rd NE, Atlanta, GA 30322\\
$^{*}$ \emph{email:} HQuick@cdc.gov
%Last updated: \today
%\end{comment}

\end{center}

\textsc{Summary.}
\begin{comment}
While death rates due to diseases of the heart have experienced a sharp decline over the past 50 years,
these diseases continue to be the leading cause of death in the United States, and the rate of decline varies by geographic location, race, and gender.
We look to harness the power of hierarchical Bayesian methods to
obtain a clearer picture of the declines from
%analyze a dataset comprised of
county-level, temporally varying heart disease death rates for men and women of different races from the US. Specifically, we propose a nonseparable multivariate spatio-temporal Bayesian model which not only allows for group-specific temporal correlations, but also allows for temporally-evolving covariance structures in the multivariate spatio-temporal component of the model. Furthermore, the model is capable of seamlessly handling counties with missing data due to the lack of observations from a particular subpopulation. After verifying the effectiveness of our model via simulation, we apply our model to a dataset of over 200,000 county-level heart disease death rates. In addition to yielding a superior fit than other common approaches for handling such data, the richness of our model provides insight into racial, gender, and geographic disparities underlying heart disease death rates in the US which are not permitted by more restrictive models. %, such as those which assume separability. {\color{red}Too long.}
\end{comment}
While death rates due to diseases of the heart have experienced a sharp decline over the past 50 years, these diseases continue to be the leading cause of death in the United States, and the rate of decline varies by geographic location, race, and gender.  We look to harness the power of hierarchical Bayesian methods to obtain a clearer picture of the declines from county-level, temporally varying heart disease death rates for men and women of different races in the US. Specifically, we propose a nonseparable multivariate spatio-temporal Bayesian model which allows for group-specific temporal correlations and temporally-evolving covariance structures in the multivariate spatio-temporal component of the model. After verifying the effectiveness of our model via simulation, we apply our model to a dataset of over 200,000 county-level heart disease death rates. In addition to yielding a superior fit than other common approaches for handling such data, the richness of our model provides insight into racial, gender, and geographic disparities underlying heart disease death rates in the US which are not permitted by more restrictive models.


\newpage

\doublespacing
\section{Introduction}
%\citet{wikle:berliner:cressie} use a space-time vector autoregressive (VAR) to model temperature data.
%\item paper 1 \citet{qbc}
{Despite substantial reductions in death rates since the mid-1960s \citep[e.g.,][]{sempos,ford:capewell,young,greenlund}, heart disease remains the leading cause of death in the United States \citep[US,][]{deaths}.}  Work by \citet{casper:changes} has identified that while the nation as a whole has experienced substantial declines in heart disease mortality rates, there has been a substantial geographic shift over time, as mortality rates in the northeast have declined at a much faster rate than those in the Deep South. Previous work has also shown disparities in heart disease death rates between the sexes {\citep[e.g.,][]{sempos,kramer:apc}}, between races {\citep[e.g.,][]{kramer:apc}}, and geographically {\citep[e.g.,][]{gillum,adam:geo,adam:methods}}, yet accounting for these various sources of disparities simultaneously has yet to be considered.  Here, we look to build upon the existing heart disease literature to obtain a broader picture of these declining death rates using a hierarchical Bayesian statistical approach which accounts for correlation spatially, temporally, and between race/gender groups.

There is an extensive literature on the subject of space-time modeling, particularly in the Bayesian context.
A common approach for modeling discrete --- or areal --- spatial data is the use of the conditionally autoregressive (CAR) model proposed by \citet{besag} and later popularized in the disease mapping context by \citet{bym}.
Early uses of the CAR model in the space-time setting {include \citet{waller:carlin} and \citet{knorr-held:besag}} ---
both of which analyzed rates of lung cancer in Ohio counties ---
%whose work relates to the field of small area disease rates,
and \citet{gelfand98}, whose interest pertained to the sale prices of homes.  While these methods have used \emph{separable} model structures
for space and time,
%which separates the spatial and temporal effects in the model structure,
\citet{knorr-held:2000} discusses the use of nonseparable space-time models in a discrete space, discrete time setting with an application to lung cancer mortality rates in Ohio.
%Oftentimes, these space-time data models can be written in the form
In addition to space-time data models, \citet{gelfand:mcar} and \citet{carlin:banerjee:2003} have developed methods for general multivariate spatial models.
%methods for developing general multivariate spatial models, such as the multivariate CAR (MCAR) models of \citet{gelfand:mcar} and \citet{carlin:banerjee:2003}. {\color{red}(not currently a sentence)}
%, are quite popular and continued to be used \citep[e.g.,][]{qbc}. {\color{red}I think you can come up with something better than this...}
For a more complete coverage of the recent advances in spatial and space-time modeling, see \citet{cressie:wikle} and \citet{BCG}.

The concept of multivariate space-time (MST) models for discrete spatial data has also been explored previously.  For instance, \citet{congdon} modeled suicide mortality rates in the boroughs of London using spatially varying regression coefficients and a nonparametric specification of the random effects, and   \citet{daniels} {developed a conditionally specified model} for the analysis of particulate matter and ozone data collected from monitoring sites in Los Angeles, CA.
%{\color{red}Add ref to meth paper.}
More recently, \citet{jon} proposed an alternative in which the authors use a shared component model \citep[e.g.,][]{knorr-held:best, tzala} with a reduced-rank spatial domain, extending the approach of \citet{hughes:haran} to the MST setting.  While a shared component model can offer substantial computational benefits by effectively reducing the complexity of a MST model to that of a reduced-rank space-time model, this assumption may not always be appropriate (e.g.,
when the available covariate information is insufficient to capture the differences in the geographic patterns)
%when the individual groups may experience dramatically different geographic patterns)
nor necessary (e.g., when number of groups in the multivariate structure is small).

Due to the recent geographic and temporal evolutions in heart disease death rates,
%evolving nature of these data,
the methodological goal of this paper is to define a nonseparable multivariate space-time modeling framework
to analyze the heart disease mortality rate data described in Section~\ref{sec:data}.
%for the analysis of the heart disease mortality rate data, which are described in Section~\ref{sec:data}.
We propose our model in Section~\ref{sec:methods} and demonstrate its ability to accurately estimate model parameters via simulation study in Section~\ref{sec:sim}.  We then analyze our heart disease mortality data in Section~\ref{sec:anal}, where we {observe temporally-evolving variance parameters inconsistent with the previously used separable model}.  Finally, we summarize our findings and offer some concluding remarks in Section~\ref{sec:disc}.

\section{Data Description}\label{sec:data}
%{\color{red}Needs to be rewritten so it's not identical to paper 1.}
The study population for this analysis includes US residents, ages 35 and older, who were identified on a death certificate as either black or white --- we restrict our analysis to these $N_g = 4$ groups because these are the only racial groups for whom data are available for the entire duration of our study period.
%{\color{red}\textbf{[Is there a particular reason why we only use blacks and whites (other than sample size issues)?]}}
Annual counts of heart disease deaths per county per race/gender group were obtained from the National Vital Statistics System (NVSS) of the National Center for Health Statistics (NCHS).
Due to differences in the manner in which
%Only a 50\% sample of
death records were processed by NCHS, % in 1972, therefore
we restrict the analysis to data from 1973--2010 to
%maximize the number of heart disease deaths in each county.
ensure valid comparisons across time.
Deaths from heart disease were defined as those for which the underlying cause of death was ``diseases of the heart'' according to the 8th, 9th, and 10th revisions of the International Classification of Diseases (ICD)\footnote{ICD--8: 390-398, 402, 404, 410--429; ICD--9: 390--398, 402, 404--429; ICD-10: I00--I09, I11, I13, I20--I51}.
Based on the works of \citet{icd8:icd9} and \citet{icd9:icd10}, we assume that this definition is consistent over the 38 year study period.
Annual projected population counts were obtained from the {\citet{census}}, %{\color{red}[bridged census data; Michele: what's the citation for this?]},
%{\color{red}and heart disease death rates per 100,000 were calculated} and
and the numbers of heart disease deaths were age-standardized to the 2000 US Standard Population using 10 year age groups.  %{These data have been previously analyzed by \citet{quick:cdc1} and \citet{adam:declines}.}\footnote{{\color{red}Drop this sentence}}

The geographic unit used in this analysis was the county (or county equivalent).  Given changes in county definitions during the study period (e.g., the creation of new counties), a single set of $N_s$ = 3,099 counties from the contiguous lower 48 states was used for the entire study period.  In an attempt to stabilize the data, county-level age-standardized counts and populations were aggregated into $N_t = 19$ two year intervals (i.e., 1973--74, 1975--76, etc.).

\section{Methods}\label{sec:methods}
\subsection{Review of methods for disease mapping}\label{sec:review}
One of the seminal papers in the field of disease mapping was the work of \citet{bym}.  Letting $Y_i$ and $n_i$ denote the incidence of disease and the population at risk in county $i$, the authors proposed a model of the form
%modeling the incidence of a disease in county $i$, $Y_i$, from a population at risk as
\begin{equation}
Y_{i} \sim \Pois\left(n_i \exp\left[\bx_i\bbeta + Z_i + \phi_i\right]\right), i=1,\ldots,N_s \label{eq:bym}
\end{equation}
where $\bx_i$ denotes a $p$-vector of covariates with corresponding regression coefficients, $\bbeta$, $Z_{i}$ is a spatial random effect, and $\phi_i \sim N\left(0,\tau^2\right)$ is an exchangeable random effect.  In their work, \citet{bym} modeled $\bZ=\left(Z_1,\ldots,Z_{N_s}\right)'$ as arising from an intrinsic conditional autoregressive (CAR) model, which has the conditional distribution
\begin{equation}
Z_i\given \bZ_{(i)},\sig^2 \sim N\left(\sum_{j=1}^{N_s} w_{ij} Z_j \slash \sum_{j=1}^{N_s} w_{ij}, \sig^2\slash \sum_{j=1}^{N_s} w_{ij}\right) \label{eq:car}
%Z_i\given \bZ_{(i)},\sig^2 \sim N\left(\sum_{j\sim i} Z_j \slash m_i, \sig^2\slash m_i\right) \label{eq:car}
\end{equation}
where $\bZ_{(i)}$ denotes the vector $\bZ$ with the $i$th element removed and $W$ is an adjacency matrix with elements $w_{ij}=1$ if $i$ and $j$ are neighbors {(denoted $i\sim j$)} and 0 otherwise.
%$j\sim i$ indicates that counties $i$ and $j$ are adjacent, and $m_i$ denotes the number of counties adjacent to the $i$th county.
Later work by \citet{knorr-held:prec} and \citet{hodges:prec} has shown that the joint distribution of $\bZ$ in~\eqref{eq:car} is of the form
\begin{equation}
\pi\left(\bZ \given \sig^2\right) \propto \left(\sig^2\right)^{-(N_s-1)\slash 2} \exp\left[-\frac{1}{2\sig^2}\bZ'(D-W)\bZ\right],\label{eq:jcar}
\end{equation}
where $D$ is an $N_s\times N_s$ diagonal matrix with elements $m_i = \sum_{j=1}^{N_s} w_{ij}$. % and $W$ is an adjacency matrix with elements $w_{ij}=1$ if $i$ and $j$ are neighbors {\color{red}(denoted $i\sim j$)} and 0 otherwise.

Extending~\eqref{eq:bym}--\eqref{eq:jcar} to a setting consisting of multiple spatial surfaces is straightforward.  For instance, suppose we wish to map $N_g$ diseases over an area consisting of $N_s$ counties.  Letting $Y_{ik}$ denote the incidence of disease $k$ in county $i$, we may assume
\begin{equation}
Y_{ik} \sim \Pois\left(n_{ik} \exp\left[\bx_{ik}\bbeta_k + Z_{ik} + \phi_{ik}\right]\right), i=1,\ldots,N_s, \;k=1,\ldots,N_g. \label{eq:bym_k}
\end{equation}
To model
$\bZ = \left(\bZ_{1\cdot}',\ldots,\bZ_{N_s\cdot}'\right)'$ where $\bZ_{i\cdot} = \left(Z_{i1},\ldots,Z_{iN_g}\right)'$,
%{\color{red}$\bZ = \left(\bZ_{\cdot1}',\ldots,\bZ_{\cdot N_g}'\right)'$ where $\bZ_{\cdot k} = \left(Z_{1k},\ldots,Z_{N_sk}\right)'$},
we may follow the example of \citet{gelfand:mcar} and let $\bZ\sim\MCAR\left(1,\Sig_Z\right)$ which yields the following: %joint distribution
\begin{align*}
\pi\left(\bZ\given \Sig_Z\right) &\propto \vert \Sig_Z\vert^{(N_s-1)\slash 2} \exp\left[-\frac{1}{2}\bZ'\left\{(D-W)\otimes\Sig_Z^{-1}\right\}\bZ\right]\\
\text{and}\;\;\bZ_{i\cdot}\given \bZ_{(i)\cdot},\Sig_Z &\sim N\left(\sum_{j\sim i} \bZ_{j\cdot} \slash m_i, \frac{1}{m_i}\Sig_Z\right),
\end{align*}
\begin{comment}
\begin{equation*}
\pi\left(\bZ\given \Sig_Z\right) \propto \vert \Sig_Z\vert^{(N_s-1)\slash 2} \exp\left[-\frac{1}{2}\bZ'\left\{(D-W)\otimes\Sig_Z^{-1}\right\}\bZ\right]
\end{equation*}
and the conditional distribution
\begin{equation*}
\bZ_{i\cdot}\given \bZ_{(i)\cdot},\Sig_Z \sim N\left(\sum_{j\sim i} \bZ_{j\cdot} \slash m_i, \frac{1}{m_i}\Sig_Z\right),
\end{equation*}
\end{comment}
where $\Sig_Z$ denotes the $N_g \times N_g$ covariance structure for our $N_g$ diseases and $\otimes$ denotes the Kronecker product.  Extensions of~\eqref{eq:bym} to the (multivariate) space-time setting follow similarly \citep[e.g.,][]{waller:meth}, with the necessary specifications of the covariance matrix $\Sig_Z$.
%or the multivariate space-time

{%\color{red}
While Poisson models like~\eqref{eq:bym_k} are common, they can also pose computational challenges.  For instance, the full conditional of $\bZ_i$, given by
\begin{align}
\pi\left(\bZ_i\given \bY,\bZ_{(i)\cdot},\bbeta,\bphi,\Sig_Z\right) \propto \prod_{k=1}^{N_g}\Pois\left(Y_{ik}\given \bbeta_k, Z_{ik}, \phi_{ik}\right) \times \pi\left(\bZ_i\given \bZ_{(i)\cdot},\Sig_Z\right)
\end{align}
is \emph{not} a known distribution.  That is, if we use a Markov chain Monte Carlo (MCMC) algorithm to estimate the posterior distribution of our model parameters, this model may require the use of Metropolis steps within our Gibbs sampler. When the number of groups is large --- or in the space-time setting when $N_t$ is large --- updating $\bZ$ can be cumbersome.
\citet{besag:poisson} suggests a reparameterization of~\eqref{eq:bym_k} which would involve integrating $\phi_{ik}$ out of the model, yielding a Gaussian full conditional for $\bZ_i$, though this model still consists of over twice as many parameters as data points.
%{\color{blue}{knorr-held:rue}} have developed
%{\color{blue}Need to explain why this is bad...}
}

One alternative to modeling the counts using a Poisson likelihood is to model the \emph{rates} as being log-normally distributed.  For instance, suppose $Y_{ikt}$ and $n_{ikt}$ denote the number of heart disease-related deaths and the population at risk for the $k$th population in county $i$ at time $t$, respectively. We could then model
%\begin{equation}
$\theta_{ikt} = \log\left(Y_{ikt}\slash n_{ikt}\right)$ %\label{eq:log}
%\end{equation}
using a Gaussian distribution. % {\citep[e.g.,][]{haining}}.
This %Unfortunately, however, this
%may not be an option in our case,
may be problematic, however,
as our data consist of a large number of counties
%in which $Y_{ikt}=0$
experiencing \emph{zero} deaths related to heart disease
for a given population in a given year. As such, this may require
%As $\log(x)$ is not defined for $x=0$, this may require
us to treat $Y_{ikt}=0$ as data below the limit of detection by substituting $Y_{ikt} = Y_{ikt}^* < 1$ or by multiply imputing values for $Y_{ikt}$ \citep[e.g., see][]{fridley:dixon}. %That said, this seems inappropriate as zero \emph{is} our observed number of deaths.
%us to instead model $\theta_{ikt}^* = \log\left(\left[Y_{ikt} + \zeta\right]\slash n_{ikt}\right)$, for some small $\zeta$. We could also reframe the issue by treating $Y_{ikt}=0$ as data below the limit of detection , $1\slash n_{ikt}$, using an approach similar to \citet{paper4}.

In order to avoid the computational burden associated with the Poisson model in~\eqref{eq:bym} and the ill-handling of zeros in the log-normal model, %in~\eqref{eq:log},
we opt to model the rates themselves as Gaussian.  That is, we let $Y_{ikt}$ denote the age-standardized death rate (per 100,000) in county $i\in\{1,\ldots,N_s\}$ during time interval $t=\{1,\ldots,N_t\}$ for race/gender group $k\in\{1,\ldots,N_g\}$ and we define
$\bY_{i\cdot t}$ to be the vector collecting the $N_g$ observations from time $t$ in the $i$th county, $\bY_{i\cdot\cdot}=\left(\bY_{i\cdot1}',\ldots,\bY_{i\cdot N_t}'\right)'$ to be the vector collecting the $\left(N_gN_t\right)$ observations from the $i$th county, and $\bY = (\bY_{1\cdot\cdot}',\ldots,\bY_{N_s\cdot\cdot}')'$ to be the $N_sN_gN_t$-vector which stacks all of the age-standardized death rates.
To model the death rates, we assume
%and assume
\begin{equation}
Y_{ikt} \sim N\left(\bx_{ikt}'\bbeta_k + Z_{ikt},\tau_{ikt}^2\right),\;i=1,\ldots,N_s,\;k=1,\ldots,N_g,\;t=1,\ldots,N_t\label{eq:Y}
\end{equation}
where $\bx_{ikt}$ is the $p\times1$ vector of covariates for the $i$th county at time $t$ with a corresponding $p\times1$ vector of regression coefficients, $\bbeta_k$, $Z_{ikt}$ is a random effect which accounts for the spatio-temporal dependence between and within the four race/gender groups, $\tau_{ikt}^2 = \tau_k^2\slash n_{ikt}$, and $n_{ikt}$ denotes the population of group $k$ in county $i$ at time $t$ divided by 100,000.
A recent example of a model of this form is \citet{qbc}, where a Gaussian likelihood was used to model changes in county-level asthma hospitalization rates in California.
{We provide a defense of the Gaussian assumption for these data in {Figure B.1} of the Web Appendix.}
%The Gaussian assumption appears valid (see the Web Appendix)}

\subsection{Choices for $\Sig_Z$}\label{sec:special}
Before we present our proposed $\MSTCAR$ model for $\Sig_Z$ in Section~\ref{sec:mstcar}, we begin by describing other natural choices: independence models and a separable model.  Not only do these models have computational benefits, but they are also special cases of the $\MSTCAR$.
\subsubsection{Independence models}
Based on the multivariate spatial models described in Section~\ref{sec:review}, one could opt to fit a collection of $N_g$ independent space-time models (denoted $\STCAR$) of the form
\begin{align}
\bZ_{ik\cdot}\given\bZ_{(i)k\cdot},\sig_k^2,\rho_k &\sim N\left(\frac{1}{m_i} \sum_{j\sim i} \bZ_{jk\cdot}, \frac{\sig_k^2}{m_i} R\left(\rho_k\right)\right), k=1,\ldots,N_g\label{eq:ind}
\end{align}
where $R\left(\rho_k\right)$ denotes an $N_t \times N_t$ temporal correlation matrix with parameter $\rho_k$ and $\sig_k^2$ is the variance parameter corresponding to race/gender group $k$.  In addition to accounting for spatiotemporal correlation, a model of the form~\eqref{eq:Y} with this structure for $\bZ$ has the added computational benefit of being able to be fit \emph{in parallel} as $N_g$ separate models.  This convenience, however, comes at the cost of failing to account for the correlation between groups.  As we believe there to be a high degree of correlation between the heart disease mortality rates of our various race/gender groups, this drawback is particularly disappointing.

We could also choose to fit $N_t$ independent multivariate spatial models of the form
\begin{align*}
\bZ_{i\cdot t}\given\bZ_{(i)\cdot t},G_t &\sim N\left(\frac{1}{m_i} \sum_{j\sim i} \bZ_{j\cdot t}, \frac{1}{m_i} G_t\right),
\end{align*}
where $G_t$ denotes a temporally-varying $N_g \times N_g$ multivariate covariance structure for our race/gender groups.  While this model can also have substantial computational benefits, the assumption of \emph{temporal} independence is especially damning. %, in that we assume \emph{temporal} independence.

\subsubsection{Separable model}
{%\color{red}
Driven by the desire to account for both temporal and between-group correlation in our spatial model, a \emph{separable} model of the form
\begin{align}
\bZ_{i\cdot\cdot}\given\bZ_{(i)\cdot\cdot},G,\rho &\sim N\left(\frac{1}{m_i} \sum_{j\sim i} \bZ_{j\cdot\cdot}, \frac{1}{m_i} R(\rho)\otimes G\right),\label{eq:sep}
\end{align}
%by letting $\Sig_Z = R(\rho)\otimes G$,
where we let $R\left(\rho\right)$ denote an $N_t \times N_t$ temporal correlation matrix and $G$ denote the $N_g \times N_g$ between-group covariance structure, may be attractive.  The appeal of a separable model where $\Sig_Z = R(\rho)\otimes G$ is immediately clear: instead of accounting for multivariate temporal correlation using an unstructured $N_gN_t \times N_gN_t$ matrix, $\Sig_Z$, we can \emph{separate} our problem into matrices of rank $N_g$ and $N_t$, reducing the computational complexity of inverting $\Sig_Z$ substantially.  While the criticism of separable models in the spatiotemporal literature is primarily directed toward their use in the continuous space, continuous time setting where prediction at unobserved locations is of interest \citep[e.g., see][]{stein2005}, the lack of a temporally evolving $G_t$ or group-specific $\rho_k$ may be {undesirable}.

%For instance, setting $\rho_k\equiv \rho$ for all $k$ and $G_t \equiv G$ for all $t$ would yield the \emph{separable} multivariate spatiotemporal CAR structure of the form: {\color{red}(Add reference for a separable model)} While the separable model in~\eqref{eq:sep} reduces computational burden by avoiding the direct inversion of $\Sig_{\eta}$ in~\eqref{eq:Zcar}, this will likely result in only a marginal improvement in computational time when $N_g$ and $N_t$ are small {\color{red}(Can I give some evidence of this somewhere?  i.e., in sim or results sections)}.  In contrast, the $\STCAR$ model in~\eqref{eq:ind} can be fit as $N_g$ parallel models, yielding a substantial improvement in computation time.%\footnote{{\color{red}This paragraph can be moved after the hierarchical model in a subsection called ``Special cases''}}
}
\subsubsection{The $\MSTCAR$ model}\label{sec:mstcar}
\begin{comment}
We let $Y_{ikt}$ denote the age-standardized death rate in county $i\in\{1,\ldots,N_s\}$ during time interval $t=\{1,\ldots,N_t\}$ for race/gender group $k\in\{1,\ldots,N_g=4\}$.
%, computed by dividing the age-standardized number of deaths by the population at risk and multiplying by {\color{red}10,000} to obtain a ``per 10,000 persons'' rate.
We define
$\bY_{i\cdot t}$ to be the vector collecting the $N_g$ observations from time $t$ in the $i$th county, $\bY_{i\cdot\cdot}=\left(\bY_{i\cdot1}',\ldots,\bY_{i\cdot N_t}'\right)'$ to be the vector collecting the $\left(N_gN_t\right)$ observations from the $i$th county, and $\bY = (\bY_{1\cdot\cdot}',\ldots,\bY_{N_s\cdot\cdot}')'$ to be the $N_sN_gN_t$-vector which stacks all of the age-standardized death rates.  To model the death rates, we assume %{\color{red}check into log rates}
\begin{equation}
Y_{ikt} \sim N\left(\bx_{ikt}'\bbeta_k + Z_{ikt},\tau_{ikt}^2\right),\;i=1,\ldots,N_s,\;k=1,\ldots,N_g,\;t=1,\ldots,N_t %\label{eq:Y}
\end{equation}
where $\bx_{ikt}$ is the $p\times1$ vector of covariates for the $i$th county at time $t$ with a corresponding $p\times1$ vector of regression coefficients, $\bbeta_k$, $Z_{ikt}$ is a random effect which accounts for the spatio-temporal dependence between and within the four race/gender groups, $\tau_{ikt}^2 = \tau_k^2\slash n_{ikt}$, and $n_{ikt}$ denotes the population of group $k$ in county $i$ at time $t$ divided by 100,000. %This modeling choice, however, is not without its potential drawbacks, which we discuss further in Section~\ref{sec:disc}.
\end{comment}

To construct our random effects, $\bZ$, we will begin by defining $\bv_{\iota\cdot t}\iid N(\bzero,G_t)$ to be a collection of independent $N_g$-dimensional random variables with covariance $G_t$ for $\iota=1,\ldots,(N_s-1)$ and $t=1,\ldots, N_t$.
Note the deliberate use of the subscript $\iota$ instead of the subscript $i$; this is to reinforce that $\bv_{\iota\cdot t}$ does not correspond to a particular county.
From this, %we can construct
we define $\bv_{\iota k\cdot} = \left(v_{\iota k1},\ldots,v_{\iota kN_t}\right)'$ and construct
\begin{align}
\bet_{\iota k\cdot} = \tR_k \bv_{\iota k\cdot} \sim N\left(\bzero,\tR_k G_{k,k}^* \tR_k'\right),\;\iota=1,\ldots,(N_s-1),\;k=1,\ldots,N_g,\label{eq:eta}
\end{align}
where we define $\tR_k$ to be the Cholesky decomposition of $R_k$ such that $\tR_k \tR_k' = R_k$, where $R_k \equiv R\left(\rho_k\right)$ is a temporal correlation matrix based on an autoregressive order 1 --- or AR(1) --- structure with correlation parameter $\rho_k$ and $G_{k,k}^*$ is the $N_t \times N_t$ diagonal matrix with elements $\left\{G_t\right\}_{k,k}$ for $t=1,\ldots,N_t$.
%This choice of temporal correlation structure is not necessary for our model, but it does permit certain computational benefits (see Web Appendix A for details).
Equivalently, we can define $\bet_{\iota\cdot\cdot} \sim N\left(\bzero,\Sig_{\eta}\right)$ where
\begin{align}\label{eq:Sig_eta}
\Sig_{\eta} =
\begin{bmatrix}
\tR_{1,1}^* & \bzero & \bzero\\
\vdots & \ddots & \bzero \\
\tR_{N_t,1}^* & \cdots & \tR_{N_t,N_t}^*
\end{bmatrix}
\begin{bmatrix}
G_1 & \bzero & \bzero\\
\bzero & \ddots & \bzero \\
\bzero & \bzero & G_{N_t}
\end{bmatrix}
\begin{bmatrix}
\tR_{1,1}^* & \cdots & \tR_{N_t,1}^*\\
\bzero & \ddots & \vdots \\
\bzero & \bzero & \tR_{N_t,N_t}^*
\end{bmatrix},
\end{align}
where $\tR_{t,t'}^*$ denotes the $N_g\times N_g$ diagonal matrix with elements $\left\{\tR_k\right\}_{t,t'}$ for $k=1,\ldots,N_g$.
Finally, we let $\Sig_Z \equiv \Sig_{\eta}$ and define $\bZ$ in the form of an $\MCAR\left(1,\Sig_{\eta}\right)$
%in a manner similar to the $\MCAR$ of
of \citet{gelfand:mcar} with
a conditional and (improper) joint distribution of
\begin{align}
\bZ_{i\cdot\cdot}\given \bZ_{(i)\cdot\cdot},G_{1},\ldots,G_{N_t},\brho \sim& \;N\left(\frac{1}{m_i} \sum_{j\sim i} \bZ_j, \frac{1}{m_i} \Sig_{\eta}\right),\;i=1,\ldots,N_s \label{eq:Zcar}\\
\pi\left(\bZ\given G_{1},\ldots,G_{N_t},\brho\right) \propto& \;\vert \Sig_{\eta} \vert^{-(N_s-1)\slash2} \exp\left[-\frac{1}{2} \bZ' \left\{(D-W) \otimes \Sig_{\eta}^{-1}\right\} \bZ\right], \label{eq:Zcar_joint}
\end{align}
\begin{comment}
a conditional distribution of
%those proposed by \citet{gelfand:mcar} and \citet{carlin:banerjee:2003}:
\begin{align}\label{eq:Zcar}
\bZ_{i\cdot\cdot}\given \bZ_{(i)\cdot\cdot},G_{1},\ldots,G_{N_t},\brho \sim N\left(\frac{1}{m_i} \sum_{j\sim i} \bZ_j, \frac{1}{m_i} \Sig_{\eta}\right),\;i=1,\ldots,N_s
\end{align}
%where $m_i$ denotes the number of neighbors for the $i$th county.
and an improper joint distribution of the form
\begin{align}\label{eq:Zcar_joint}
\pi\left(\bZ\given G_{1},\ldots,G_{N_t},\brho\right) \propto& \vert \Sig_{\eta} \vert^{-(N_s-1)\slash2} \exp\left[-\frac{1}{2} \bZ' \left\{(D-W) \otimes \Sig_{\eta}^{-1}\right\} \bZ\right].
\end{align}
%where $\otimes$ denotes the Kronecker product.
%Note that the $N_s-1$ in~(\ref{eq:eta}) is due to the use of the intrinsic CAR model in~\eqref{eq:Zcar} and~\eqref{eq:Zcar_joint}.
We denote the expression in~(\ref{eq:Zcar_joint}) as $\bZ \sim \MSTCAR\left(G_{1},\ldots,G_{N_t},\brho\right)$.
%{\color{red}Note that a number of commonly used models can be written as a special case of the $\MSTCAR$.  }
\end{comment}
respectively.  We denote the expression in~(\ref{eq:Zcar_joint}) as $\bZ \sim \MSTCAR\left(G_{1},\ldots,G_{N_t},\brho\right)$.

\subsection{Hierarchical model}
We complete the hierarchial model by specifying prior distributions for the remaining parameters.  As is common in Bayesian modeling, we place a flat, noninformative prior on $\bbeta_k$, and, following {\citet{gelman2006}}, assume an improper uniform prior over the positive real numbers for $\tau_k$.  For each of the spatio-temporal covariance matrices, $G_t$, we assume an inverse Wishart distribution with positive definite scale matrix $G$ and $\nu>N_g-1$ degrees of freedom,
%when $N_s$ is small and additional structure is needed, one can put a hyperprior on the matrix $G$, allowing the model to borrow strength across the $G_t$.
and we use Beta priors for each of the $\rho_k$s.
Finally, as many rural counties (particularly in the north-central states) have no data from the black populations, we decompose $\bY$ as $\bY_c = \left(\bY_o',\bY_u'\right)'$, where $\bY_o$ denotes the vector of counties with observed populations and $\bY_u$ denotes the vector of counties with unobserved populations.  Putting these pieces together, the full hierarchical model is as follows:
\begin{align}\label{eq:hier}
\pi\left(\bbeta,\bZ,G_1,\ldots,G_t,\brho,\btau^2,\bY_u\given\bY_o\right) \propto& N\left(\bY\given X\bbeta+\bZ,\Sig_Y\right)
\times \MSTCAR\left(\bZ\given G_{1},\ldots,G_{N_t},\brho\right)\notag\\
&\times \prod_{k=1}^{N_g} \left[\tbeta\left(a_{\rho},b_{\rho}\right) \times \pi\left(\tau_k^2\right)\right] \notag\\
&\times \prod_{t=1}^{N_t} \IW\left(G_t\given G,\nu\right),
\end{align}
where the notation $\pi(x)$ denotes the marginal distribution for a random variable $x$ and $\pi(x\given y)$ denotes the conditional distribution of $x$ given $y$.  Here, $\Sig_Y$ is a diagonal matrix with elements $\tau_{ikt}^2$, $X$ is the $(N_sN_gN_t\times p)$ matrix of covariates,
%$\bZ = \left(\bZ_{1\cdot\cdot}',\ldots,\bZ_{N_g\cdot\cdot}'\right)'$, $MCAR\left(\bZ\given G,\rho\right)$ denotes the joint distribution (a multivariate CAR model) induced by~(\ref{eq:Zcar}),
and $\pi\left(\tau_k^2\right)$ is the density for $\tau_k^2$ which corresponds to a flat prior for $\tau_k$. In cases where it may be difficult to learn about each $G_t$ or each $\rho_k$, %{\color{red}(what situations?)},
we may consider putting additional structure on the priors for these parameters. Note that in~(\ref{eq:hier}), $\bY_u$ is treated as an unknown model parameter, and thus each $Y_{ikt}\in\bY_u$ is sampled from~(\ref{eq:Y}) during each iteration of the MCMC algorithm. Furthermore, we assign a small value for each $n_{ikt}$ in the set $\{n_{ikt}: Y_{ikt}\in\bY_u\}$. A detailed derivation of the MCMC sampler used for this analysis, as well as a description of the benefits of using an AR(1) model to account for temporal correlation, can be found in Web Appendix~A.

\section{Simulation Study}\label{sec:sim}
To evaluate the ability of our model to accurately estimate all of our model parameters, we devised two simulation studies, each comprised of $L=100$ sets of data generated using our $\MSTCAR$ model with $N_t=10$ timepoints, $N_g=3$ groups, and the $N_s=58$ counties of California as our spatial domain.  This spatial domain offered a compromise between creating a computationally feasible simulation study (compared to using all 3,099 county equivalents) while representing a state with a moderate number of counties and variation in population density and geographic spread.  The first simulation study assumes that $n_{ikt}\equiv n$ for all combinations of $(i,k,t)$, allowing us to focus on parameter estimation irrespective of the amount of information each county can provide.  We will then relax this assumption by generating data using actual populations of California counties.

In each simulation study, performance was primarily assessed via coverage (i.e., the percent of 95\% credible intervals (95\% CI) which cover the true parameter values) where values near 95\% are desired.  Furthermore, we will compare results from the $\MSTCAR$ model proposed here to {those obtained using a separable model}. While the separable model will be incapable of providing accurate estimates for the many additional parameters which comprise $\Sig_{\eta}$, the focus here will be on model fit.  Specifically, we will compare the coverage of $\bZ$ and the deviance information criterion (DIC) of \citet{dic}, where lower values indicate a better compromise of model fit and model complexity. %\footnote{\color{red}We should add the separable model into this simulation study... does it yield similar coverage for $\bZ$?  If so, this is important to note (i.e., that separable models can still capture the lack of separability, they simply fail to provide you the tools to determine that that is occurring)}

\subsection{Equal population sizes}\label{sec:sim1}
The $\ell$th dataset is created by generating $Y_{ijk}^{(\ell)} \sim N(Z_{ijk}^{(\ell)},\tau_k^2)$ where $\tau_k^2=1$ for $k=1,\ldots,N_g$ and $\bZ^{(\ell)}$ is drawn from the $\MSTCAR$ model in~(\ref{eq:Zcar_joint}).  To do this, we first let $\brho = (0.8, 0.85, 0.90)'$ and generated samples of $G_t$ from an inverse Wishart distribution with {$2*N_g+1$ degrees of freedom} and {scale matrix $20*N_g *I_{N_g}$}, where $I_{N_g}$ is the identity matrix of size $N_g$.  Using these parameters to construct $\Sig_{\eta}$ (from which all $L$ datasets are based), we generated our latent variables $\bet_{\iota\cdot\cdot}^{(\ell)} \sim N\left(\bzero,\Sig_{\eta}\right)$.  From these, we used the methods described in \citet{rue:held} to generate our $\bZ^{(\ell)}$; specifically, we found the eigenvalues and eigenvectors of the matrix $D-W$ (based on the adjacencies of counties in California) and used the linear dependence of the eigenvectors to generate our spatial structure.  Each simulated dataset is then analyzed using the hierarchical model in~(\ref{eq:hier}) using MCMC.  Using the priors described in the previous section, % and {\color{red}$G_t \sim InvWish(30*I_{N_g},N_g+.1)$}\footnote{{\color{red}For G70, you may want to rerun this simulation using a little bit better prior... just saying}},
we initialized all of our parameters (including $\bZ$) at their true values, resulting in chains which were quick to converge and allowing us to assess the performance of our model using just 1,500 iterations of our MCMC algorithm, the last 500 of which were used as the basis for our results.  In order to better visualize these results, we also display results from an arbitrarily selected dataset.

\begin{figure}[t]
    \centering
        \includegraphics[width=.95\textwidth]{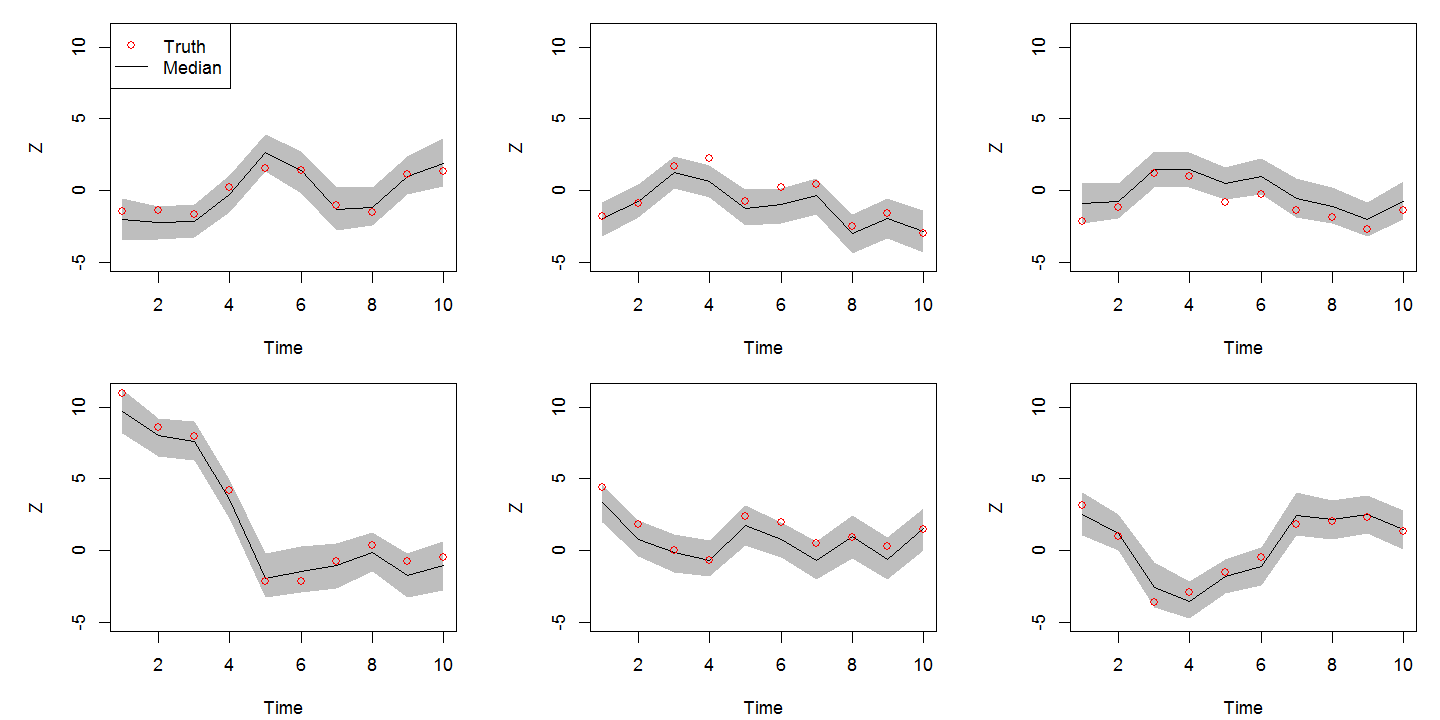}
    %\end{center}
    \caption{Selected $\bZ_{ik\cdot}$ curves from one dataset of the first simulation study.  Plots in the same row correspond to the same county, and plots in the same column correspond to the same group.  Black lines denote posterior medians, red circles denote true values, and gray bands denote the 95\% CI.} %{\color{red}Change everything to ``per 100,000''}}
    \label{fig:sim_Z}
\end{figure}

\begin{figure}[t]
    \centering
        \includegraphics[width=.95\textwidth]{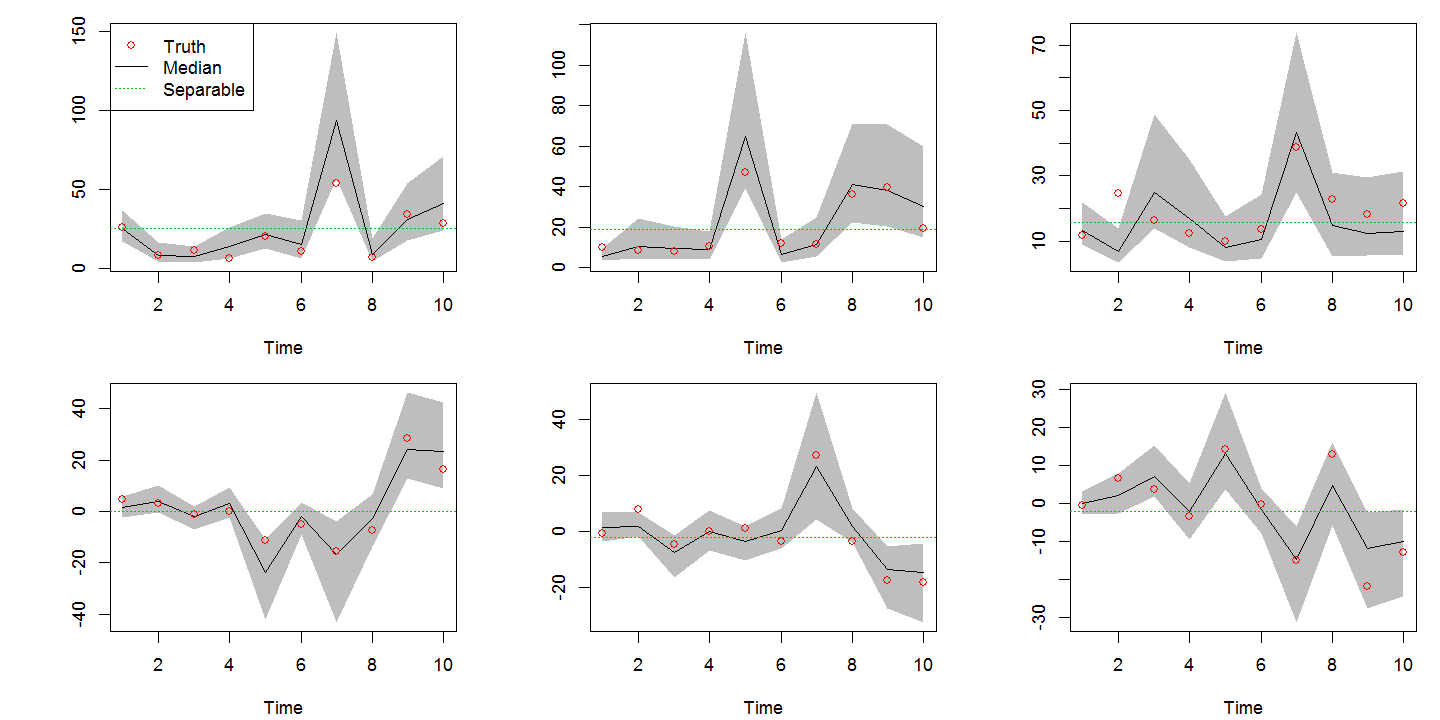}
    %\end{center}
    \caption{Estimated $G_{t;k,k'}$ from one dataset of the first simulation study.  Top row displays the diagonal elements, while the bottom row displays the off-diagonal elements.  Black lines denote posterior medians, red circles denote true values, and gray bands denote the 95\% CI.  For comparison purposes, the green lines denote the analogous values from the separable model.} %{\color{red}Change everything to ``per 100,000''}}
    \label{fig:sim_G}
\end{figure}

Overall, our model performed quite well.  Collectively, the $Z_{ikt}$ were well estimated, as demonstrated in Figure~\ref{fig:sim_Z}, with our model obtaining 94.4\% coverage and offering an improvement in DIC in 82 of the 100 datasets.  This accuracy is permitted due in part to the flexibility of our model to allow for temporally evolving $G_t$.  As shown in Figure~\ref{fig:sim_G}, the randomly generated $G_t$ exhibited some irregular behavior.  While the $\MSTCAR$ model was able to estimate these $G_t$ quite well --- with 95.4\% coverage for the diagonal elements (i.e., the variances) and 95.3\% coverage for the off-diagonal elements (i.e., the covariances) --- the separable model fails to accommodate such a nuanced multivariate structure.
%, which ---unlike the separable model, can allow for sudden, large jumps in the $Z_{ikt}$.  Fortunately, the model was able to estimate these $G_t$ quite well, with 95.4\% coverage for the diagonal elements (i.e., the variances) and 95.3\% coverage for the off-diagonal elements (i.e., the covariances).
We also achieved accurate estimates of the error variances, $\tau_k^2$, for which we obtained an average of 91.3\% coverage. In contrast, coverage for $\rho_k$ was less than ideal (85\%).

\subsection{Varying population sizes}\label{sec:sim2}
In our second simulation study, we generated data using the same design as described in Section~\ref{sec:sim1}, but here we assigned $n_{ikt}$ to be the population of the $i$th county at time $t$ for the following subpopulations: white men ($k=1$), white women ($k=2$), and black men and women ($k=3$).  While
white men and women have
%the first two groups have
$n_{ikt}>200$ in all counties for all time periods,
%our third group has
there are many counties with small black population sizes.  As such, we combine black men and women to limit the number of counties with no data.
%, including one county with \emph{no} population during one time period.
In cases where a county has {no} population during time $t$, however, we assume $n_{ikt}=1$ and treat $Y_{ikt}$ as missing.

Our model was again able to obtain accurate estimates for the $Z_{ikt}$ and the various elements of the $G_t$ while outperforming the separable model in all 100 datasets (based on DIC).  Furthermore --- aside from an expected increase in the width of the credible intervals --- there does not appear to be any degradation in the estimation for these parameters as we shift from the well-populated groups to the third, less populated group.  Unfortunately,
our model again performs less well with respect to the temporal correlation parameters, $\rho_k$.  It is understandable, though, how the problem from our first example would be exacerbated here, as the amount of information provided by each group depends on the county populations. % sizes.

\subsection{General findings}\label{sec:sim_disc}
In both simulation studies, the $\MSTCAR$ was able to obtain accurate estimates of both the $Z_{ikt}$ and the $G_t$.  While the nonseparable model offered improved DIC when compared to the separable model, it is important to note that the differences were not substantial, with just over a 1\% reduction on average. This suggests that the key benefit of the $\MSTCAR$ model (with respect to model fit) is that it provides more precise results (i.e., narrower credible intervals) than the separable model while still achieving the desired coverage.

Based on these results, %trace and autocorrelation plots\footnote{\color{red}Show in Appendix?},
the $\rho_k$ parameters appear to be difficult to identify. % and may require many more samples in order to obtain accurate estimates.
As such, if inference on the $\rho_k$ is desired, it may be necessary to run our MCMC algorithms for more iterations and consider \emph{thinning} our samples to obtain samples which are less correlated over the course of the chain.  Another option would be to consider respecifying our priors for the $\rho_k$.  In these simulation studies, we had assumed a $\tbeta(9,1)$ prior for $\rho_k$, but a more informative prior may be appropriate, particularly in the case of varying population sizes. For instance, we could assume a multi-level model of the form
$%\begin{align*}
\rho_k \sim \tbeta\left(\upsilon_{\rho}\rho_0,\upsilon_{\rho}(1-\rho_0)\right)$ for $k=1,\ldots,N_g$, %\;\text{and}\;\rho_0 \sim Beta\left(a_0,b_0\right),
%\end{align*}
where $\rho_0 \sim \tbeta\left(a_0,b_0\right)$ and $\upsilon_{\rho}$ is a parameter which controls the informativeness of the prior.  In extreme cases, we may even consider forcing $\rho_k \equiv \rho_0$, which can be induced by letting $\upsilon_{\rho}\to\infty$.  In addition to improving the convergence of our MCMC algorithm, this may also lead to minor computational benefits while still yielding a model that is more flexible than the separable model in~\eqref{eq:sep}.
%{\color{red}Is there anything we could do to help identify $\rho_k$?}

\section{Analysis of Heart Disease Death Rates}\label{sec:anal}
We fitted the nonseparable hierarchical model in~(\ref{eq:hier}) to the heart disease mortality data described in Section~\ref{sec:data} using covariates consisting of only an intercept term for each combination of 2-year time-interval and race/gender \citep[as required, per][]{besag95}, forcing the random effects to account for a substantial amount of the spatio-temporal variability in the data.
%Based on previous work with similar data \citep[e.g.,][who analyzed data from the total US population]{casper:changes},
We place a $\tbeta(9,1)$ prior on each of the $\rho_k$ to encourage higher temporal correlations in the model,
%{\color{red}Lance asks if we could use a more informative prior}),
and we use a vague inverse Wishart prior for each of the $G_t$.
We ran the MCMC algorithm with a single chain for {6,000} iterations, diagnosing convergence via trace plots for many of the model parameters and discarding the first 1,000 iterations as burn-in.
% {\color{red}(might need evidence of this in supplement)}.
Following that, we thinned our posterior samples by removing 9 out of 10 samples --- while this is not theoretically necessary, it {reduced the burden of storing excess samples for our over 200,000 random effects}. Estimates provided are based on posterior medians, and 95\% credible intervals (95\% CI) were obtained by taking the 2.5- and 97.5-percentiles from the thinned post-burn-in samples. To determine if the %computational
burden associated with fitting this nonseparable model was necessary, we compared our model %'s performance
to the separable model in~\eqref{eq:sep} and the $N_g$ independent $\STCAR$ models in~\eqref{eq:ind}. % using DIC. %\footnote{\color{red}Easy fix here}

Table~\ref{tab:dic} displays the results of our model comparison.  Here, it is clear that the independent $\STCAR$ models --- while computationally convenient --- are inadequate for these data, as both the separable and $\MSTCAR$ models offer improvements in DIC of over {94,000 units}.  As seen in Section~\ref{sec:sim}, the separable and $\MSTCAR$ models appear to perform similarly, with the $\MSTCAR$ model having a DIC only 5,828 units lower.  Given the evidence in the literature that DIC tends to favor over-fitted models \citep[e.g.,][]{robert:dic:disc}, it remains unclear if the flexibility of the $\MSTCAR$ model is \emph{required} here; nevertheless, we will henceforth focus our attention on results from the $\MSTCAR$ model.
%{\color{red}Lance asks if we can show a scatterplot of means from separable and nonseparable to see where the differences are}

\begin{table}[t] %t...
\normalsize
%\begin{doublespace}
\begin{center}
\begin{tabular}{|l|c|c|}
\hline
Model & DIC & $p_D$\\
\hline
$\STCAR$ & 2,423,049 & 32,110\\
Separable & 2,334,355 & 24,185\\
$\MSTCAR$ & 2,328,527 & 25,699\\
\hline
\end{tabular}
\end{center}
\caption{{Model fit comparison between the independent $\STCAR$ models, a separable model, and the nonseparable $\MSTCAR$ model proposed here.  Lower values of DIC indicate a better compromise of model fit and model complexity, where $p_D$ is a measure of model complexity.}}
\label{tab:dic}
%\end{doublespace}
\end{table}

\begin{figure}[t]
    \centering
        \includegraphics[width=.6\textwidth]{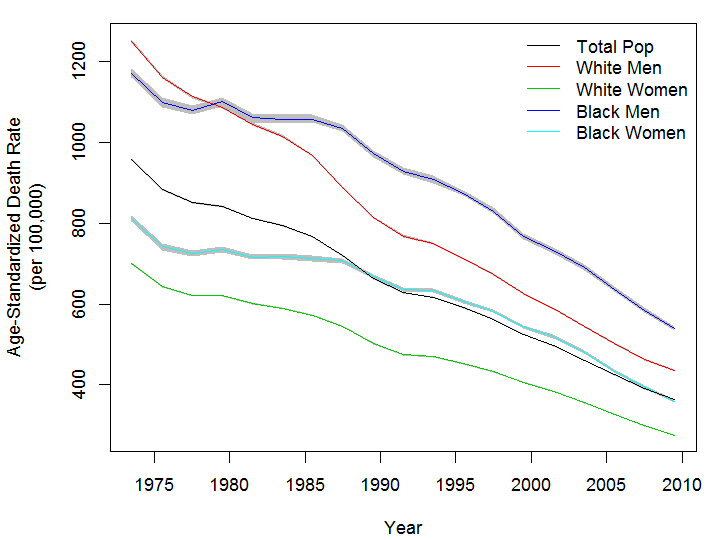}
    %\end{center}
    \caption{Heart disease death rates over time for each of the race/gender groups compared to the total population.  Gray bands denote the 95\% credible intervals for the estimates. } %{\color{red}Change everything to ``per 100,000''}}
    \label{fig:trend}
\end{figure}

Figure~\ref{fig:trend} displays the expected nationwide death rate trends for each group.  These trend lines were computed by first computing the posterior distribution for the expected value for $Y_{ikt}$ as
%\begin{equation*}
$\widehat{Y}_{ikt} = \bx_{ikt}'\bbeta_k + Z_{ikt}$.
%\end{equation*}
We then estimated the nationwide death rate for group $k$ at time $t$ by constructing the posterior for
\begin{equation*}
\widehat{Y}_{\cdot kt} = \frac{\sum_i \widehat{Y}_{ikt} n_{ikt}}{\sum_i n_{ikt}}.
\end{equation*}
A number of important findings can be found from this figure.  First and foremost, all four of our race/gender groups have experienced substantial declines, with death rates being more than cut in half.  Secondly, men of both races experience significantly higher rate of heart disease-related death than women. That said, men and women of both races do not decline at the same rate; e.g., while white men began the study as the population with the highest risk, they were soon surpassed by black men, whose rates appear to be relatively stagnant for the period from 1975--76 to 1987--88.  This trend is also visible for black women.

\begin{sidewaysfigure}[t]
    \begin{center}
        \subfigure[1973--74]{\includegraphics[width=.4\textwidth]{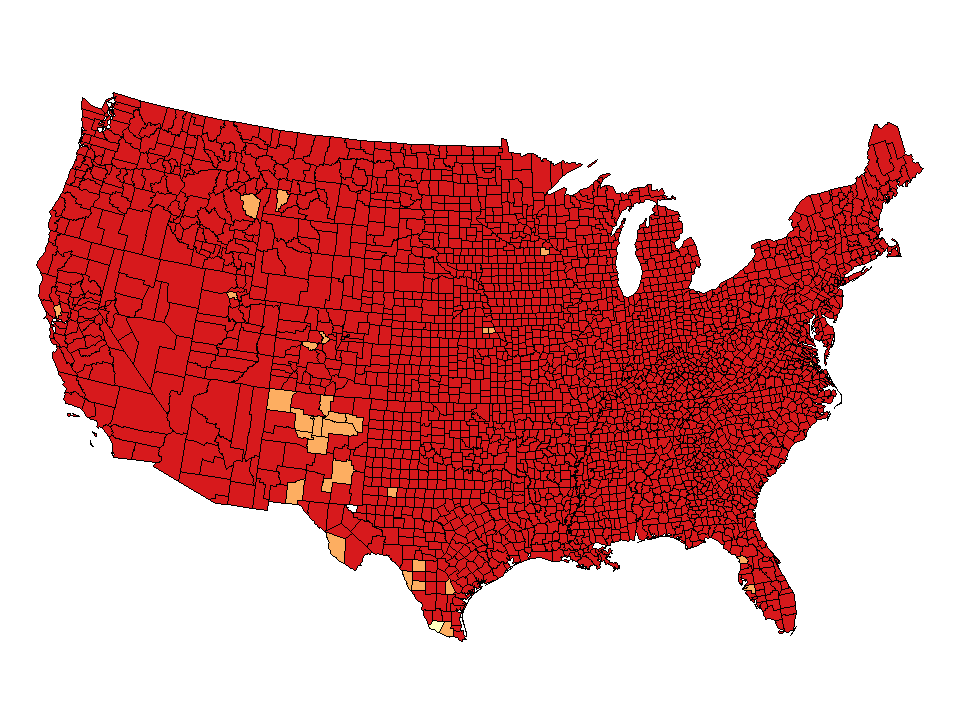}\label{fig:wm1}}
        \subfigure[1985--86]{\includegraphics[width=.4\textwidth]{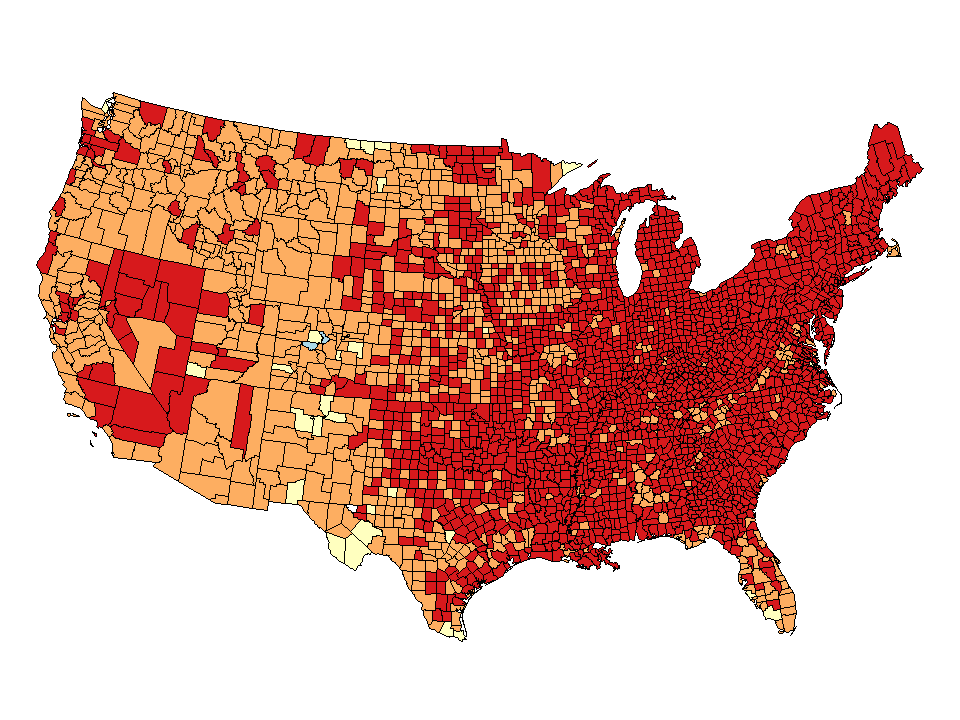}\label{fig:wm2}}
        \hspace*{.9in}\\
        \subfigure[1997--98]{\includegraphics[width=.4\textwidth]{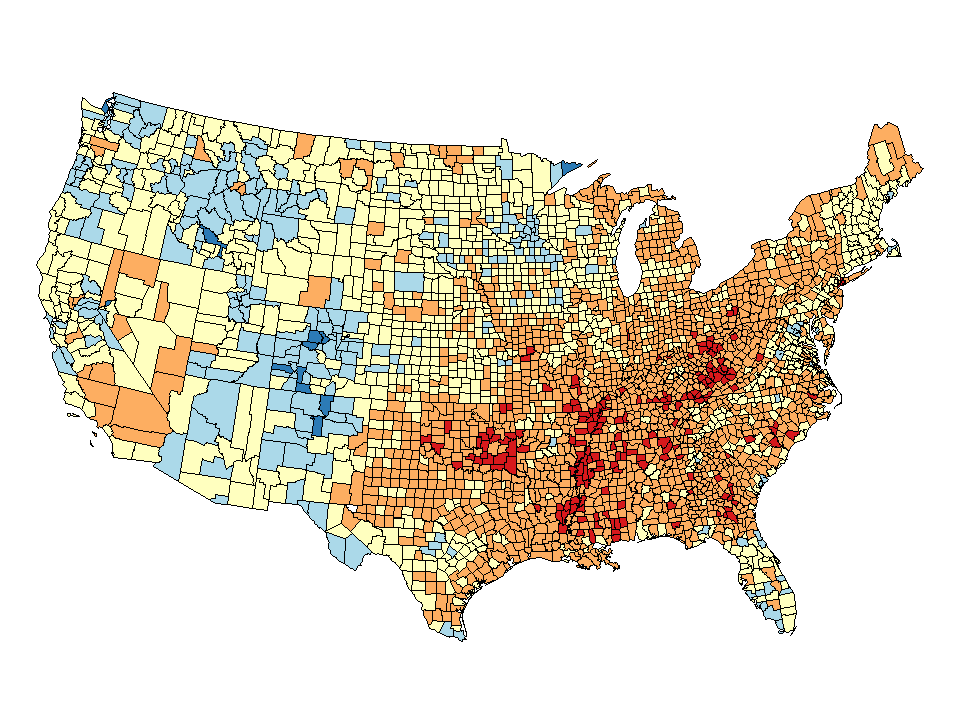}\label{fig:wm3}}
        \subfigure[2009--10]{\includegraphics[width=.4\textwidth]{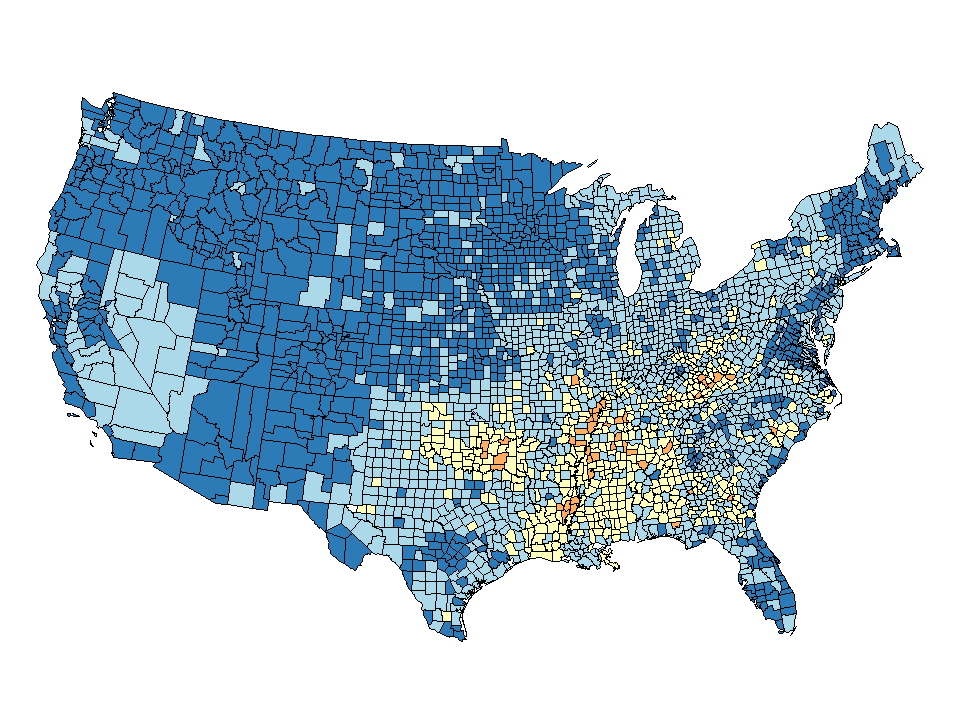}\label{fig:wm4}}
        \includegraphics[width=.04\textwidth]{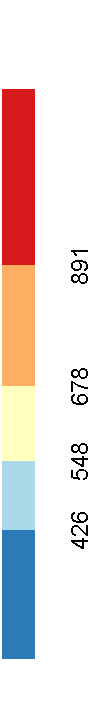}
    \end{center}
    \caption{Estimated expected heart disease death rates (per 100,000) for white men for selected time-intervals.  %Color cut-offs range from blue to red, with cuts at 426, 548, 678, and 891 deaths per 100,000.
    }
    \label{fig:white_men}
\end{sidewaysfigure}

To illustrate the changing geographic patterns, Figure~\ref{fig:white_men} displays heart disease death rates for white men for four time-intervals.  Here, we notice an interesting trend, as several major cities (e.g., Denver, CO; Washington, DC; Atlanta, GA; Minneapolis, MN) %; San Antonio, TX)
are consistently leading the charge toward lower rates of heart disease related death for white men in their respective regions.  On the other hand, there are
collections %a number of clusters
of counties in which rates are lagging behind, most prominently along the southern %south of Missouri along the
Mississippi River and much of the Deep South. Similar patterns can be found for the remaining race/gender groups. %, as shown in Figures~B.4 and~B.5 of the Web Appendix.
%{Web Appendix~B}.

%\begin{itemize}
%    \item Denver; MPLS; San Antonio / Austin; Madison, WI; Washington, DC; Boston, MA; Atlanta, GA
%    \item Kentucky/WV;
%\end{itemize}

\begin{figure}[t]
    \begin{center}
        \subfigure[White Men]{\includegraphics[width=.3166\textwidth]{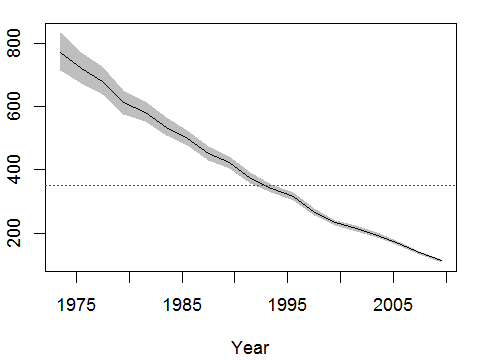}\label{fig:Sigwm}}
        \subfigure[White Women]{\includegraphics[width=.3166\textwidth]{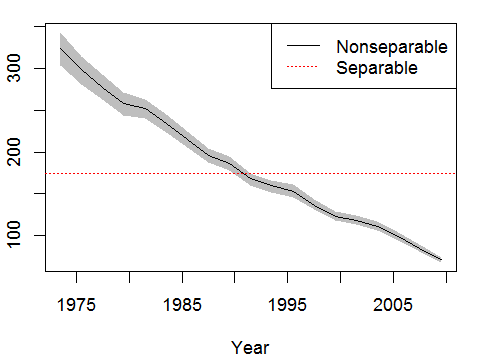}\label{fig:Sigwf}}\\
        \subfigure[Black Men]{\includegraphics[width=.3166\textwidth]{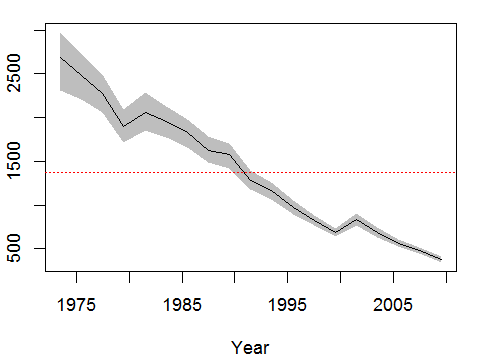}\label{fig:Sigbm}}
        \subfigure[Black Women]{\includegraphics[width=.3166\textwidth]{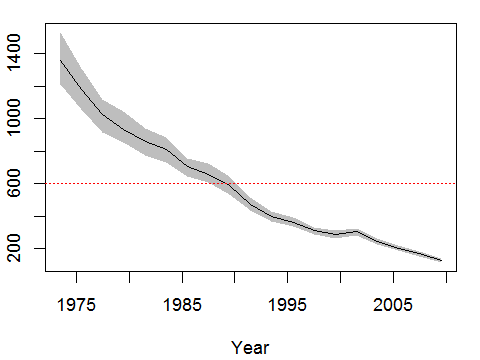}\label{fig:Sigbf}}
    \end{center}
    \caption{Estimated diagonal elements of $\Sig_{\eta}$ for the heart disease mortality data for our four race/gender groups.  Black lines denote posterior medians from the nonseparable model while gray bands denote 95\% CI.  For comparison purposes, the red line indicates the analogous value from the separable model which is constant over time.}
    \label{fig:heart_Sig}
\end{figure}

We now turn our attention to the numerous variance parameters permitted by the use of the nonseparable model.  While in Section~\ref{sec:sim} we presented posterior distributions for the elements of $G_t$ (Figure~\ref{fig:sim_G}), these parameters are not necessarily of direct interest as they are the variance parameters for $\bv_{\ell\cdot t}$, and thus they are \emph{not} directly interpretable on the scale of the data.  Instead, we need to use our posterior samples of $G_t$ and $\rho_k$ to construct $\Sig_{\eta}$ from~\eqref{eq:Sig_eta}.  These values coincide to the conditional covariance matrix of $\bZ_{i\cdot\cdot}$ (when scaled by the number of neighbors, $m_i$), and thus \emph{are} {interpretable on the scale of the data}.  Figure~\ref{fig:heart_Sig} displays the diagonal elements of $\Sig_{\eta}$ from the nonseparable model, as compared to the analogous estimates from the separable model.  Here, we find --- for all race/gender groups --- that the variability of $Z_{ikt}$ has decreased substantially from the beginning of the study period to the end.  More importantly, however, we note that the separable model severely underestimates the variance at the beginning of the study and severely overestimates the variance at the end.  {As shown in Figure~B.2 of the Web Appendix}, this can lead to oversmoothing when the rates are the highest (the 1970s) and undersmoothing when the rates are lowest (the 2000s), neither of which is desirable.  {This may be due to the fact that the rates themselves decline over time.}
%While not shown here, c
Correlations between race/gender groups are all non-zero, with high correlations %($>0.9$)
between genders of the same race and moderate correlations %($\approx 0.60$)
between races; these results can be found in {Figure B.3} of the Web Appendix.

\section{Discussion}\label{sec:disc}
In this paper, we have proposed a nonseparable framework for the purpose of modeling a dataset comprised of temporally-varying county-level heart disease death rates for multiple race/gender populations.  We evaluated the validity of the proposed methodology --- referred to as the $\MSTCAR$ model --- via simulation and demonstrated that the model was capable of providing a good fit to the data and obtaining accurate estimates for the many variance parameters. {Not only did the $\MSTCAR$ model outperform two more conventional models, but we show our model can help control the degree of smoothing in data which undergo a substantial temporal evolution during the study period.}

\begin{comment}
Extensions:
\begin{itemize}
    \item For manageable $N_s$, one could envision models with $G_{it}$, $\rho_{ik}$, and/or $\alpha_{kt}$, where $\alpha_{kt}$ denotes the CAR propriety parameter (assumed to be 1 in the improper CAR structures used here).
    \item For cases where $N_s$ is large (like this one), one may also consider using dimension reduction techniques such as those proposed by \citet{hughes:haran} and implemented by \citet{jon}.  Unfortunately, it's unclear whether or not this would actually result in computational gains, as the approach of \citet{hughes:haran} destroys the conditional properties which make CAR models attractive.  That is, when implementing the $\MSTCAR$ model proposed here, one need only invert and manipulate matrices of dimension $N_gN_t$ to sample the $\bZ_{i\cdot\cdot}$, albeit this requires looping through each of the $N_s$ areal regions.  The approach of \citet{hughes:haran}, however, replaces this $N_s$ loop with a single $N_s^*N_gN_t$-dimensional update, where $N_s^*\ll N_s$ is the rank of the reduced spatial domain.  Were we to reduce the dimension of our spatial domain from $N_s=3,099$ to $N_s^*=310$ (a 90\% reduction), this would now require manipulating $N_s^*N_gN_t = 23,560$-dimensional matrix, which would not be feasible in our setting. {\color{red}Technically, I think you could at least reduce it to $N_s^*N_g$ matrices, but this is still $1200+$.}
\end{itemize}
\end{comment}

While the methods proposed here are much more sophisticated than more commonplace models like those discussed in Section~\ref{sec:special}, %\footnote{{\color{red}If you remove references to your separable paper, you need to tweak this.}}
there are a number of extensions which could be used to enhance the $\MSTCAR$ model.  For manageable values of $N_s$, for instance, one could envision models with region-specific parameters $G_{it}$ and $\rho_{ik}$.  Implementing these models would likely require the use of a proper CAR model (e.g., the model proposed in Section~\ref{sec:methods} is constructed using only $N_s-1$ latent vectors), say by replacing $(D-W)$ in~(\ref{eq:Zcar_joint}) with $(D-\alpha_{kt}W)$, where $\alpha_{kt}\in[0,1)$ ensures propriety and $\alpha_{kt}=1$ yields the improper CAR-based model used here.  Furthermore, one may choose to use a multi-level modeling approach for specifying priors for many of these parameters, such as
%\begin{align*}
%G_{it} &\sim InvWish\left(\nu_i G_t,\nu_i\right)\\
%G_{t} &\sim Wish\left(\nu G_0,\nu\right)\\
%G_{0} &\sim InvWish\left(\nu_0 G,\nu_0\right)
%\end{align*}
\begin{align*}
G_{it} \sim \IW\left(\nu_i G_t,\nu_i\right),\;G_{t} \sim \Wish\left(1\slash\nu G_0,\nu\right),\;\text{and}\;G_{0} &\sim \IW\left(\nu_0 G,\nu_0\right)
\end{align*}
to facilitate additional borrowing-of-strength. {Computational burden and identifiability concerns notwithstanding}, such a model would be rather intuitive to specify and construct; i.e., one could let $\bet_{i\cdot\cdot} \sim N\left(\bzero,\Sig_{\eta_i}\right)$, where $\Sig_{\eta_i}$ is constructed as in~(\ref{eq:Sig_eta}) with $i$ subscripts.  Based on the results of \citet{hcar} --- where the authors extended a separable space-time model to allow for region-specific variance parameters --- there is evidence to believe that models of this sort may offer substantial improvements in fit.

For cases where $N_s$ is large, one may also consider using dimension reduction techniques such as those proposed by \citet{hughes:haran} and extended by \citet{jon}.  Unfortunately, it's unclear whether or not this would actually result in computational gains in our setting without making additional assumptions, as the approach of \citet{hughes:haran} removes the conditional properties which make CAR models attractive.  That is, when implementing the $\MSTCAR$ model proposed here, one need only invert and manipulate matrices of rank $N_gN_t$ to sample the $\bZ_{i\cdot\cdot}$, albeit this requires looping through each of the $N_s$ areal regions.  An analogous approach based on \citet{hughes:haran}, however, would replace this $N_s$ loop with a single $N_s^*N_gN_t$-dimensional update, where $N_s^*\ll N_s$ is the rank of the reduced spatial domain.  Were we to reduce the dimension of our spatial domain from $N_s=3,099$ to $N_s^*=310$ (a 90\% reduction), this would still require manipulating matrices of rank $N_s^*N_gN_t = 23,560$, which would not be feasible in our setting. While one could take advantage of the AR(1) structure to ease the burden, this would result in the manipulation of matrices of rank $N_s^*N_g$, which may \emph{still} be too large to implement in practice without resorting to the shared component model of \citet{jon}.

In the immediate future, we have two primary areas for next steps.  Motivated by this and earlier work, we aim to investigate the observed geographic disparities in heart disease death rates by identifying potential factors which may be associated with the patterns observed here.
%our primary goal
In addition to further exploring the mechanics driving heart disease death rates, we plan to apply a similar modeling framework to data comprised of county-level \emph{stroke}-related death rates.  As stroke data are typically more erratic with much lower rates of incidence, these data will present additional challenges.  In particular, the normal approximation used in this analysis will be less appropriate; as such, we aim to explore the possibility of implementing this methodology in a log-linear modeling framework using a Poisson likelihood.

%\citet{census} something \citet{census}
\begin{comment}
\section*{Stuff to add}
\begin{itemize}
    \item Need to add at least one more recent MCAR example
\end{itemize}
\end{comment}

\bibliographystyle{jasa}
\bibliography{cdc_ref,cdc_epi}
%\bibliography{cdc_epi}

\end{document}